\documentclass[twocolumn,showpacs,preprintnumbers,amsmath,amssymb]{revtex4}

\usepackage{graphicx}

\usepackage{dcolumn}

\usepackage{bm}

\newcommand{\ket}[1]{|#1\rangle}

\begin{document}

\title{Dephasing of a superconducting qubit induced by photon noise}

\author{P. Bertet$^1$, I. Chiorescu$^{1*}$, G. Burkard$^{2,3}$, K. Semba$^{1,4}$, C. J. P. M. Harmans$^1$, D. P. DiVincenzo$^{2}$, J. E. Mooij$^1$}

\affiliation{$^1$Quantum Transport Group, Kavli Institute of Nanoscience, Delft University of Technology, Lorentzweg $1$, $2628CJ$, Delft, The Netherlands \\
$^2$ IBM T.J. Watson Research Center, P.O. Box 218, Yorktown
Heights, NY 10598, USA \\
$^3$ Department of Physics and Astronomy, University of Basel,
Klingelbergstrasse 82, CH-4056 Basel, Switzerland \\
$^4$ NTT Basic Research Laboratories, Atsugi-shi, Kanagawa
243-0198, Japan \\
$^*$Present address : National High Magnetic Field Laboratory,
Florida State University, 1800 East Paul Dirac Drive Tallahassee,
Florida 32310, USA.}

\begin{abstract}

We have studied the dephasing of a superconducting flux-qubit
coupled to a DC-SQUID based oscillator. By varying the bias
conditions of both circuits we were able to tune their effective
coupling strength. This allowed us to measure the effect of such a
controllable and well-characterized environment on the qubit
coherence. We can quantitatively account for our data with a
simple model in which thermal fluctuations of the photon number in
the oscillator are the limiting factor. In particular, we observe
a strong reduction of the dephasing rate whenever the coupling is
tuned to zero. At the optimal point we find a large spin-echo
decay time of $4 \mu s$.

\end{abstract}


\maketitle

Retaining quantum coherence is a central requirement in quantum
information processing. Solid-state qubits, including
superconducting ones \cite{qubits,Vion02,Chiorescu03}, couple to
environmental degrees of freedom that potentially lead to
dephasing. This dephasing is commonly associated with
low-frequency noise \cite{decoherence}. However, resonant modes at
higher frequencies are harmful as well. In resonance with the
qubit transition they favor energy relaxation. Off-resonance they
may cause pure dephasing, due to fluctuations of the photon number
stored in the oscillator. Experimentally we show that the quantum
coherence of our superconducting flux-qubit coupled to a DC-SQUID
oscillator is limited by the oscillator thermal photon noise. By
tuning the qubit and SQUID bias conditions we can suppress the
influence of photon noise, and we measure a strong enhancement of
the spin-echo decay time from about $100ns$ to $4 \mu s$.

In our experiment, a flux-qubit of energy splitting $h \nu_q$ is
coupled to a harmonic oscillator of frequency $\nu_p$ which
consists of a DC-SQUID and a shunt capacitor
\cite{Chiorescu04,Bertet_resact}. The oscillator is weakly damped
with a rate $\kappa$ and is detuned from the qubit frequency. In
this dispersive regime, the presence of $n$ photons in the
oscillator induces a qubit frequency shift following
$\nu_{q,n}-\nu_{q,0}=n \delta \nu_0$, where the shift per photon
$\delta \nu_0$ depends on the effective oscillator-qubit coupling.
Any fluctuation in $n$ thus causes dephasing. Taking the
oscillator to be thermally excited at a temperature $T$ and
assuming the pure dephasing time $\tau_\phi>>1/\kappa$, we find
\cite{Bertet_condmat} after a reasoning similar to \cite{Blais} :

\begin{equation}\label{eq:tauphi}
  \tau_\phi = \frac{\kappa}{ \bar{n} (\bar{n}+1) (2 \pi \delta
\nu_0)^2}
\end{equation}

\noindent with the average photon number stored in the oscillator
$ \bar{n}=(\exp{h \nu_p /k T}-1)^{-1}$. We note that a similar
effect was observed in a recent experiment on a charge qubit
coupled to a slightly detuned waveguide resonator
\cite{Wallraffac}. When driving the oscillator to perform the
readout, the authors observed a shift and a broadening of the
qubit resonance due to the AC-Stark shift and to photon shot
noise, well-known in atomic cavity quantum electrodynamics
\cite{chat}. In our experiments, the oscillator is not driven but
thermally excited. In addition, we are able to tune in-situ the
coupling constant and $\delta \nu_0$, and therefore to directly
monitor the decohering effect of the circuit.

\begin{figure}
\resizebox{.45\textwidth}{!}{\includegraphics{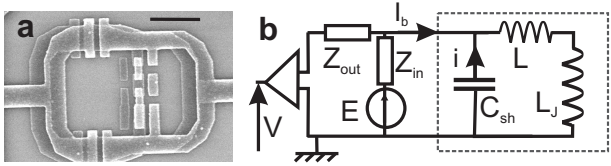}}
\caption{(a) SEM picture of the sample. The flux qubit is the
small loop containing four Josephson junctions in a row ; the
SQUID is constituted by the outer loop containing the two large
junctions. The bar equals $1 \mu m$. (b) Measuring circuit
diagram. The SQUID, represented by its Josephson inductance $L_J$,
is shunted by an on-chip capacitor $C_{sh}$ through
superconducting lines of inductance $L$, forming the plasma mode.
 \label{fig1}}
\end{figure}

Our flux-qubit consists of a micron-size superconducting aluminum
loop intersected by four Josephson junctions
\cite{Mooij99,Caspar00} fabricated by standard electron-beam
lithography and shadow evaporation techniques (see figure
\ref{fig1}a ; note that compared to earlier designs
\cite{Chiorescu03}, we added a fourth junction to restore the
qubit-SQUID coupling symmetry \cite{Guido04}). When the magnetic
flux threading the loop $\Phi_x$ sets the total phase across the
junctions $\gamma_q$ close to $\pi$, the loop has two low-energy
eigenstates, ground state $\ket{0}$ and excited state $\ket{1}$
\cite{Caspar00,Chiorescu03}. The flux-qubit is characterized by
the minimum energy separation $h \Delta$ between $\ket{0}$ and
$\ket{1}$, and the persistent current $I_p$ \cite{Mooij99}. In the
basis of the energy eigenstates at the bias point $\gamma_q=\pi$,
the qubit hamiltonian reads $H_q=-(h/2)(\Delta \sigma_z + \epsilon
\sigma_x)$, where $\epsilon \equiv (I_p/e) (\gamma_q-\pi)/(2\pi)$.
The energy separation is $E_1-E_0 \equiv h \nu_q = h \sqrt{\Delta
^2 + \epsilon ^2 }$. Note that $d \nu_q / d \epsilon = 0$ when the
qubit is biased at $\epsilon=0$ so that it is to first order
insensitive to noise in $\epsilon$, in particular to noise in the
flux $\Phi_x$. This is similar to the doubly optimal point
demonstrated in the quantronium experiment \cite{Vion02}.

The qubit is inductively coupled to a SQUID detector with a mutual
inductance $M$ (large loop in figure \ref{fig1}a), and to an
on-chip antenna allowing us to apply microwave pulses. The readout
scheme and the experimental setup have been described elsewhere
\cite{Chiorescu03}. The average persistent current in the qubit
loop, with a sign depending on its state $\ket{k}$ ($k=0,1$),
generates a flux which modifies its critical current $I_C \sim 1
\mu A$ to a value $I_C^{\ket{k}}$~; a bias current pulse of
amplitude $I_m$ chosen so that $I_C^{\ket{0}}<I_m<I_C^{\ket{1}}$
allows us to discriminate between the two states by detecting the
switching of the SQUID. Before the measurement, when the bias
current $I_b<I_C$, the SQUID behaves as a Josephson inductance
$L_J$ which depends on the flux threading it and on $I_b$. It is
connected to an on-chip capacitor $C_{sh}$ through a line with a
stray inductance $L$ (see figure \ref{fig1}b) and thus forms a
harmonic oscillator of frequency $\nu_p=1/2\pi
\sqrt{(L+L_J)C_{sh}}$ called the plasma mode
\cite{Chiorescu04,Bertet_resact} (note that the junction
capacitance is much smaller than $C_{sh}$). We can write its
hamiltonian $H_p=h \nu_p a^\dag a$, where $a$ ($a^\dag$) is the
annihilation (creation) operator. The total current flowing
through the SQUID is thus $I_b + i$, with $i= \delta i_0
(a+a^\dag)$ being the operator for the current in the plasma mode
and $\delta i_0$ the rms fluctuations of the current in the
oscillator ground state $\delta i_0=\sqrt{h \nu_p/2(L+L_J)}$. The
SQUID circuit is connected to the output voltage of our waveform
generator $E$ via an impedance $Z_{in}$, and to the input of a
room-temperature amplifier through $Z_{out}$ which define the
oscillator quality factor $Q=2 \pi \nu_p/\kappa$. $Z_{in}$ and
$Z_{out}$ take into account low-temperature low-pass filters
\cite{Chiorescu03}, and on-chip $8 k \Omega$ thin-film gold
resistors thermalized by massive heat-sinks. The resulting
impedance seen from the plasma mode is estimated to be $9k \Omega$
at low frequencies and of order $500\Omega$ at $GHz$ frequencies.
The measurements were performed at a base temperature $T_b=30mK$.

The applied magnetic field and the bias current $I_b$ result in a
circulating current $J$ in the SQUID loop \cite{Lefèvre-Seguin}.
Via the qubit-SQUID coupling $M$ the qubit phase $\gamma_q$ will
be affected, so that we can write the qubit energy bias as a sum
of two contributions $\epsilon=\eta+\lambda$, where $\eta=2 I_p
(\Phi_x-\Phi_0/2)/h$ is controlled by $\Phi_x$ and $\lambda=2 I_p
M J(I_b)/h$ only depends on $I_b$ \cite{noteonJ}. This dependence
has two important consequences. First, the qubit bias point is
shifted by the measurement pulse, allowing us to operate the qubit
at the flux-noise insensitive point while keeping a measurable
signal \cite{Chiorescu03}. Second, it gives rise to a coupling
between the qubit and the plasma mode described by a hamiltonian
$H_I=h[g_1(I_b) (a+a^\dag) + g_2(I_b) (a+a^\dag)^2] \sigma_x$,
where $g_1(I_b)=(1/2)(d \lambda / d I_b)\delta i_0$, and
$g_2(I_b)=(1/4)(d^2 \lambda / d I_b^2) (\delta i_0)^2$
\cite{Bertet_condmat}. We note that this coupling hamiltonian
depends on $I_b$ via $g_1$ and $g_2$ and is thus tunable in-situ.
In particular it is possible to cancel $g_1$ by biasing the SQUID
at a current $I_b^*$ such that $d \lambda /dI_b=0$. The qubit is
then effectively {\it decoupled} from its measuring circuit
\cite{Guido04}. Our design therefore allows us to study the effect
of the coupling between the qubit and its measuring circuit by
varying $I_b$, while keeping all other parameters unchanged.

\begin{figure}
\resizebox{.45\textwidth}{!}{\includegraphics{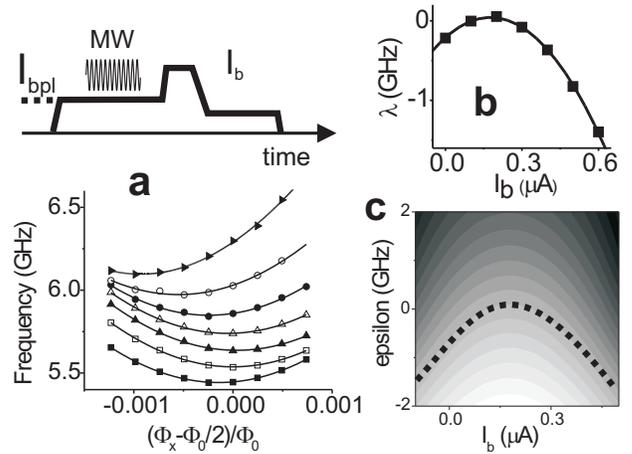}}
\caption{(a) Top : Principle of the spectroscopy experiments : a
bias current pulse of amplitude $I_{bpl}<I_C \sim 1 \mu A$ is
applied while a microwave pulse (MW) probes the qubit resonance
frequency. The qubit state is finally measured by a short bias
current pulse as discussed in \cite{Chiorescu03}. Bottom : Qubit
spectroscopy for $I_{bpl}$ varying between $0\mu A$ to $0.6\mu A$
with steps of $0.1 \mu A$ (bottom to top). The curves were offset
by $100MHz$ for clarity. The solid curves are fits to the formula
for $\nu_q$. (b) Curve $\lambda(I_b)$ deduced from the
spectroscopy curves as explained in the text. The solid line is a
parabolic fit to the data. The decoupling condition is satisfied
at $I_b^*=180 \pm 20 nA$. (c) Calculated frequency shift $\delta
\nu_0(I_b,\epsilon)$ for the parameters of our sample. The white
scale corresponds to $-20MHz$, the black to $+40MHz$. Along the
dotted line $\epsilon_m(I_b)$, $\delta \nu_0=0$.}
 \label{fig2}
\end{figure}

To obtain the coupling constants $g_1$ and $g_2$, we performed
extensive spectroscopic measurements of the qubit, as a function
of both $I_b$ and $\Phi_x$. We applied a pre-bias current pulse
$I_{bpl}$ through the SQUID while sending a long microwave pulse,
followed by a regular measurement pulse \cite{Chiorescu03} at a
value $I_m$ (see figure \ref{fig2}a). We measured the SQUID
switching probability as a function of the microwave frequency,
and recorded the position of the qubit resonance as a function of
$I_{bpl}$ and $\Phi_x$. The data are shown in figure \ref{fig2}a
for various values of $I_{bpl}$. We observe that for each bias
current, a specific value of external flux $\Phi_x^0(I_{bpl})$
realizes the optimal point condition. Fitting all the curves with
the formula $\nu_q=\sqrt{\Delta^2+[\lambda (I_{bpl})+2I_p
(\Phi_x-\Phi_0/2)/h]^2}$, we obtain the qubit parameters
$M=6.5pH$, $\Delta=5.5GHz$, $I_p=240nA$, and also $\lambda(I_b)$
which is shown in figure \ref{fig2}b together with a parabolic
fit. Decoupling occurs at $I_b^*=180 \pm 20 nA$ and not at $I_b=0$
because of a $4\%$ asymmetry of the SQUID junctions. We also
measured the parameters of the SQUID oscillator by performing
resonant activation measurements and fitting the dependence of the
resonant activation peak as a function of $I_b$ and $\Phi_x$
\cite{Bertet_resact}. We found a maximum plasma frequency
$\nu_p=3.17GHz$, $C_{sh}=7.5 \pm 2 pF$ and $L=100 \pm 20 pH$,
consistent with design values. The width of the peak also gives us
an estimate for the oscillator quality factor, $Q = 120 \pm 30$.

From the previous measurements we know the parameters of the total
hamiltonian $H=H_q+H_p+H_I$ and we can deduce the value of $\delta
\nu_0$ by second-order perturbation theory \cite{Bertet_condmat} :
$\delta \nu_0=4 [(g_1(I_b) \sin
\theta)^2\frac{\nu_q}{\nu_q^2-\nu_p^2}- g_2(I_b) \cos \theta]$
 where $\theta$ is the mixing angle, defined by $\cos
\theta=\epsilon/\sqrt{\epsilon^2+\Delta^2}$. The first term in
$\delta \nu_0$ is the usual AC-Zeeman shift obtained without using
the rotating wave approximation which is not valid in our case.
Note in particular that the sign of this shift only depends on the
sign of $\nu_q-\nu_p$ which in our experiments is always positive.
The second term is due to the dependence of the SQUID Josephson
inductance on the qubit state \cite{Adrian,Bertet_resact}, and it
has the same sign as $\epsilon$ since $g_2$ is negative.
Therefore, for some value $\epsilon_m(I_b)<0$, one obtains $\delta
\nu_0=0$. This is shown as a dashed line in figure \ref{fig2}c in
which we plot $\delta \nu_0(\epsilon,I_b)$. If dephasing is indeed
limited by thermal fluctuations, we expect the dephasing time to
be maximal along $\epsilon_m(I_b)$. We note that the curve
includes $(I_b=I_b^*,\epsilon=0)$, so that this bias point is {\it
optimal} with regard to bias current noise, flux noise and photon
noise.

\begin{figure}
\resizebox{.42\textwidth}{!}{\includegraphics{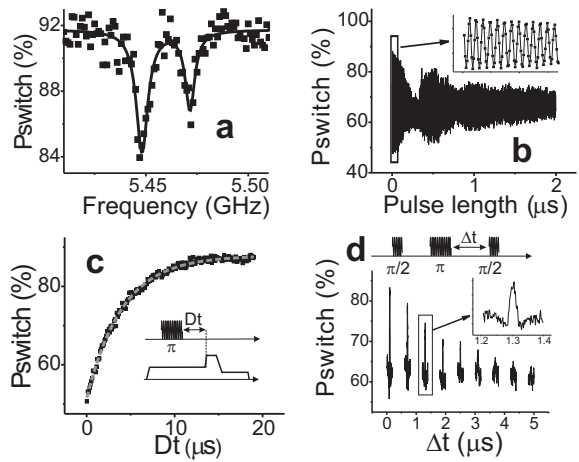}}
\caption{(a) Qubit line shape at the optimal point. The solid line
is a fit assuming a double lorentzian shape. (b) Rabi oscillations
(frequency $100MHz$) at the optimal point. The inset shows
well-behaved oscillations with nearly no damping during the first
$100ns$. (c) Measurement of $T_1$ at the optimal point ; the
dashed grey line is an exponential fit of a time constant $4\mu
s$. (d) Spin-echo pulse sequence and signal at the optimal point.
 \label{fig3}}
\end{figure}

We now turn to the measurements of the qubit coherence properties
around this optimal point, as characterized by the relaxation time
$T_1$, the qubit spectral line shape, and the spin-echo decay time
$T_{echo}$ \cite{echo}. The line shape was measured using a long
microwave pulse ($2 \mu s$) at a power well below saturation.
Figure \ref{fig3}a shows a typical result at the optimal point.
For this specific sample, we observed a twin peak structure which
likely results from one strongly coupled microscopic fluctuator.
In addition, the width of the line as well as the average value of
the gap $\Delta$ changed significantly in time, which indicates
that the residual linewidth is probably due to a larger number of
fluctuators more weakly coupled. We stress that we observed the
splitting all along the $\nu_q(\Phi_x)$ spectrum in contrast to
\cite{Simmonds}. Fitting the peaks to a sum of two Lorentzians of
widths $w_1$ and $w_2$ we define an effective dephasing time
$t_2=2/\pi (w_1+w_2)$. At the optimal point $t_2$ varied between
$50$ and $200ns$.

\begin{figure}
\resizebox{.42\textwidth}{!}{\includegraphics{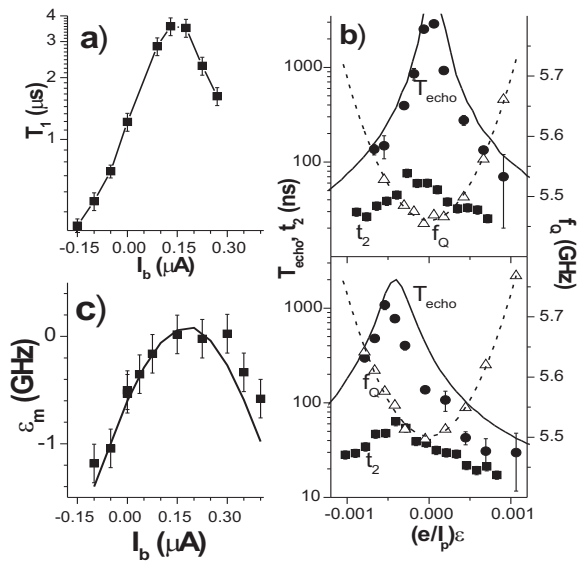}}
\caption{(a) Measurement of $T_1$ versus $I_b$ at the flux-noise
insensitive point $\epsilon=0$. (b) Measurement of $T_{echo}$
(circles), $t_2$ (squares) and of the qubit frequency (triangles),
as a function of $\epsilon$ for $I_b=I_b^*$ (top) and $I_b=0\mu A$
(bottom). The dotted line is a fit to the formula for $\nu_q$ ;
the solid black line is the prediction of equation \ref{eq:tauphi}
for $T=70mK$ and $Q=150$. (c) Value of $\epsilon$ for which $t_2$
is maximum (full squares) compared to the theoretical
$\epsilon_m(I_b)$ (full line).
 \label{fig4}}
\end{figure}

Despite the fluctuators, we were able to induce Rabi oscillations
by applying microwave pulses at the middle frequency of the split
line. An example is shown in figure \ref{fig3}b at the optimal
point. The oscillations decay non-exponentially and display a
clear beating. Nevertheless, by driving the qubit strongly enough,
we could observe well-behaved oscillations for hundreds of
nanoseconds (see inset of figure \ref{fig3}b). We measured the
energy relaxation time $T_1$ by applying a $\pi$ pulse followed
after a delay $Dt$ by a measurement pulse (see figure
\ref{fig3}c). At the optimal point, we found that $T_1=4 \mu s$.
To quantify the dephasing further we also applied the spin-echo
sequence \cite{echo}, depicted in figure \ref{fig4}a. Spin-echo
measurements are particularly relevant for our purpose, because
the photon noise in the plasma mode occurs at a relatively high
frequency set by $\kappa \simeq 130MHz$. In such conditions, this
noise affects the spin-echo damping time $T_{echo}$ as strongly as
Ramsey experiments so that $T_{echo}$ is also given by formula
\ref{eq:tauphi} \cite{Bertet_condmat} ; on the other hand the
spin-echo experiment is not sensitive to the low-frequency noise
responsible for the qubit line splitting. The results are shown in
figure \ref{fig3}d at the optimal point, by a set of curves
obtained at different delays $\Delta t$ between the $\pi$ pulse
and the last $\pi/2$ pulses. Fitting the decay of the echo
amplitude as a function of the delay between the two $\pi/2$
pulses with an exponential, we find $T_{echo}=3.9 \pm 0.1 \mu s$.
Compared with previous experiments on flux-qubits
\cite{Chiorescu03}, the long Rabi and spin-echo times were
obtained by reducing the mutual inductance $M$, and biasing the
qubit at the optimal point.

We studied the variation of $T_1$ as a function of the bias
current at the flux-insensitive point $\epsilon=0$. This required
us to adjust the flux at the value $\Phi_x^0(I_b)$. Results are
shown in figure \ref{fig4}a. We observed a clear maximum of $T_1$
for $I_b=I_b^*$. This demonstrates that at least part of the qubit
relaxation occurs by dissipation in the measuring circuit. We then
investigated the dependence of $T_{echo}$ and $t_2$ on $\epsilon$
for $I_b=I_b^*$ (figure \ref{fig4}b top, full circles and full
squares). As expected, we observe a sharp maximum for $T_{echo}$
at $\epsilon=0$ and a shallow one for $t_2$. However, at a
different bias current $I_b=0\,\mu A$, the maximum of $T_{echo}$
and $t_2$ is clearly shifted towards $\epsilon<0$. We measured the
position of this maximum in $t_2$ as a function of $I_b$ as shown
in solid squares in figure \ref{fig4}c. For dephasing caused by
flux noise or bias current noise, the maximal coherence time
should always be obtained at $\epsilon=0$ ; the observed deviation
proves that a different noise source is active in our experiments.
We find that thermally induced photon number fluctuations in the
plasma mode explains our results. In figure \ref{fig4}c we draw
the curve $\epsilon_m(I_b)$ of figure \ref{fig2}d, where the
photon-induced shift $\delta \nu_0$ equals $0$ (solid line). The
agreement between the data points and this curve, obtained
directly from measured parameters, is excellent. In addition,
assuming a reasonable effective oscillator temperature of $T=70mK$
\cite{Chiorescu04} and a quality factor of $Q=150$, which yields a
mean photon number $\bar{n}=0.15$, the dephasing time $\tau_\phi$
predicted by equation \ref{eq:tauphi} closely matches the
spin-echo measurements both for $I_b=I_b^*$ and $I_b=0\, \mu A$
\cite{noteonT1} (see the solid line in figure \ref{fig4}b). We
stress that even at such small $\bar{n}$ the photon number
fluctuations can strongly limit the qubit coherence. This suggests
that increasing the plasma frequency could lead to significant
improvement.

In conclusion, we present experimental evidence that the dephasing
times measured in a flux-qubit can be limited by thermal
fluctuations of the photon number in the SQUID detector plasma
mode to which it is strongly coupled. By careful tuning of flux
and current bias, we could decouple the qubit from its detector
and reach long relaxation and spin-echo damping times ($4 \mu s$).
These results indicate that long coherence times can be achieved
with flux qubits.

We thank Y. Nakamura, D. Est\`{e}ve, D. Vion, M. Grifoni for
fruitful discussions. This work was supported by the Dutch
Foundation for Fundamental Research on Matter (FOM), the E.U.
Marie Curie and SQUBIT grants, and the U.S. Army Research Office.


\begin{thebibliography}{99}
\bibitem{noteonJ}
In the conditions of our experiments, we can safely neglect the
additional dependence of $J$ and thus $\lambda$ on $\Phi_x$.
\bibitem{noteonT1}
We actually plot $(\tau_\phi^{-1}~+~(2T_1)^{-1})^{-1}$ where $T_1$
is the value measured at each bias current ($T_1=4\mu s$ for
$I_b=I_b^*$, and $T_1=1 \mu s$ for $I_b=0\mu A$)
\end{thebibliography}
\end{document}